\documentclass[showpacs,aps,amsmath,amssymb,twocolumn]{revtex4}
\usepackage[latin1]{inputenc}
\usepackage{graphicx}
\usepackage{bm}
\usepackage{xcolor}
\usepackage{array}
\usepackage{braket}
\usepackage{microtype}
\usepackage{wasysym}
\renewcommand{\baselinestretch}{1.0}
\begin{document}

\renewcommand{\baselinestretch}{1.0}
\title{Time-rescaled quantum dynamics as a shortcut to adiabaticity}

\author{Bert\'{u}lio de Lima Bernardo$^{1,2}$}

\affiliation{$^{1}$Departamento de F\'{\i}sica, Universidade Federal da Para\'{\i}ba, 58051-900 Jo\~ao Pessoa, PB, Brazil\\
$^{2}$Departamento de F\'{\i}sica, Universidade Federal de Campina Grande, Caixa Postal 10071, 58109-970 Campina Grande-PB, Brazil}

\email{bertulio.fisica@gmail.com}

\begin{abstract}

The design of quantum control methods has been shown to greatly improve the performance of many evolving quantum technologies. To this end, the usage of adiabatic dynamics to drive quantum systems is seriously limited by the action of environment-induced noise and decoherence. In this spirit, fast quantum processes known as shortcuts to adiabaticity have been developed as alternatives to adiabatic protocols with a myriad of potential applications.  Here, we develop a new state-independent mechanism to speed up the evolution of an arbitrary quantum dynamical system by simply rescaling the time of a reference driving process; an approach which can also work as a shortcut to adiabaticity. Our findings are illustrated for three systems, namely the parametric oscillator, the transport of a particle in a harmonic trap, and the spin-1/2 in a magnetic field.

\end{abstract}

\maketitle


\section{Introduction}

The very act of controlling the dynamics of quantum systems has continuously changed its status from being an obscure dream in the first years of quantum theory to an indispensable tool in many evolving research areas \cite{dong,con,vand,frank,senko,jurc,zhou,chu,sayrin,bert}. From a practical point of view, adiabatic quantum control techniques, which by definition are slow processes, has been shown to be particularly useful to drive and prepare states for usage in quantum computation \cite{albash}, as well as in atomic and molecular physics \cite{vit,berg}. The advantage of such techniques relies on the controllability and robustness of the system against certain types of experimental imperfections. However, on the other hand, decoherence effects and nonadiabatic leakage are more prominent in this case due to the long time evolution. To overcome these drawbacks, fast processes which reproduce the results of adiabatic ones have been proposed, the so-called ``shortcuts to adiabaticity'' (STA) \cite{torr,adc,deng}, and extensively applied to control the quantum dynamics
of single qubit in cold atoms \cite{bason}, trapped ions \cite{an}, nitrogen-vacancy centers \cite{zhou}, and superconducting qubits \cite{wang}.

 The STA protocols also offer promising applications in the emerging field of quantum thermodynamics. In this field, real quantum heat engines are expected to operate in a finite time cyclic process, a fact which, in general, gives rise to a trade-off between efficiency and power \cite{binder}. Then, one of the main challenges in the area is the optimization of the efficiency of microscopic thermal machines, while sacrificing the minimum of output power \cite{aps}. Therefore, the STA techniques come into play to reveal alternative finite time thermodynamic transformations that mimic adiabatic processes. As a matter of fact, the use of STA methods has curiously gone beyond the scope of quantum technologies to find applications in optics and classical systems \cite{odelin}. With these motivations, many techniques have been developed. For example, using dynamical invariants \cite{chen}, the fast-forward (FF) technique \cite{masuda,masuda2,torr2}, the inversion of scaling laws \cite{campo}, and the counterdiabatic driving (CD) \cite{demi,demi2,berry,deff}.

Within this class of STA scenarios, CD, also known as transitionless quantum driving, is the one that allows applications in a variety of quantum systems, as long as the spectral structure is accessible \cite{campbell,wu,saberi}. In this technique, one has initially a reference time-dependent Hamiltonian $\hat{H}_{0}(t)$, with instantaneous eigenvalues $\{E_{n}(t)\}$ and eigenkets $\{\ket{n_{t}}\}$, and from it constructs an auxiliary Hamiltonian $\hat{H}_{1}(t)$, such that their collective effect, $\hat{H}(t) = \hat{H}_{0}(t) + \hat{H}_{1}(t)$, drives the system exactly through the manifold generated by $\hat{H}_{0}(t)$ in a shorter time. Therefore, when $\hat{H}_{0}(t)$ generates an adiabatic evolution, $\hat{H}(t)$ represents an STA. It can be demonstrated that $\hat{H}_{1}(t) = i \hbar \sum_{n}(\ket{\partial_{t} n_{t}}\bra{n_{t}} - \braket{n_{t}|\partial_{t} n_{t}} \ket{n_{t}}\bra{n_{t}})$, which shows that calculating $\hat{H}(t)$ demands the instantaneous eigenkets $\ket{n_{t}}$. However, the task of obtaining these eigenkets is usually very complicated, which has limited the usefulness of the method \cite{campbell2}.                

In this work, we introduce a new state-independent scheme to speed up an arbitrary quantum process, which is taken as the reference protocol, by simply rescaling the time dependence of the Hamiltonian. Similar to the CD case, when the reference is adiabatic, the method works as a STA protocol. After establishing the general theory, our findings are illustrated for three experimentally relevant systems, namely the parametric harmonic oscillator, the transport of a particle in a harmonic trap, and the spin-1/2 in a magnetic field. Despite not being transitionless as the CD method, the construction of our fast protocol does not require knowledge about the spectrum of the system.    

\section{Time-rescaling method}

 Consider a closed quantum system on which we desire to perform a protocol according to a unitary time evolution operator $\hat{\mathcal{U}}(t,0)$, acting between an initial time $0$ and a final time $t_{f}$, with a time-dependent Hamiltonian $\hat{H}(t)$. This operator must satisfy the Schr\"{o}dinger equation,
\begin{equation}
\label{1}
\hat{H}(t)\hat{\mathcal{U}}(t,0)=i \hbar \frac{\partial}{\partial t} \hat{\mathcal{U}}(t,0).
\end{equation}
The solution of Eq.~(\ref{1}) for the case in which the Hamiltonian is time-dependent but the $\hat{H}$'s commute at different times, subject to the initial condition $\hat{\mathcal{U}}(0,0) = \mathcal{I}$, where $\mathcal{I}$ is the identity operator, is \cite{sakurai} 
\begin{equation}
\label{2}
\hat{\mathcal{U}}(t_{f},0)= \exp \left\{ -\frac{i}{\hbar} \int_{0}^{t_{f}} \hat{H}(t) dt \right\}.
\end{equation}
Here, we call it the reference evolution operator. At this point, if we rescale the time using the function $t=f(\tau)$, the above equation can be rewritten as 
\begin{align}
\label{3}
\hat{\mathcal{U}}(t_{f},0) &= \exp \left\{ -\frac{i}{\hbar} \int_{f^{-1}(0)}^{f^{-1}(t_{f})} \hat{H}(f(\tau))f'(\tau) d\tau \right\} \nonumber \\
&= \exp \left\{ -\frac{i}{\hbar} \int_{f^{-1}(0)}^{f^{-1}(t_{f})} \hat{\mathcal{H}}(\tau) d\tau \right\},
\end{align}
where $\hat{\mathcal{H}}(\tau) = \hat{H}(f(\tau))f'(\tau)$ is the time-rescaled (TR) Hamiltonian, with $f'(\tau)$ and $f^{-1}(\tau)$ being the first derivative and the inverse of $f(\tau)$, respectively. Let us call the operator of Eq.~(\ref{3}) the TR evolution operator. From Eqs.~(\ref{2}) and~(\ref{3}) we observe that, when applied to an arbitrary initial state $\ket{\psi(0)}$, the reference and TR evolutions produce exactly the same final state, $\ket{\psi(t_{f})}=\hat{\mathcal{U}}(t_{f},0) \ket{\psi(0)}$. However, this equivalence is achieved only if the reference and TR Hamiltonians, $\hat{H}$ and $\hat{\mathcal{H}}$, are applied between the corresponding time (integration) intervals. In the latter case, the desired  action of $\hat{\mathcal{H}}$ materializes in the time interval between $\tau = f^{-1}(0)$ and $\tau = f^{-1}(t_{f})$.

Let us clarify in more detail the importance in the freedom of writing the time evolution operator $\hat{\mathcal{U}}(t_{f},0)$ of Eq.~(\ref{2}) in the form shown in Eq~(\ref{3}). As already indicated, in both equations the resulting effect of $\hat{\mathcal{U}}(t_{f},0)$ is precisely the same, independent of the initial state $\ket{\psi(0)}$ on which they act. On the one hand, when written in the first (reference) form, the time evolution is generated by the Hamiltonian $\hat{H}$ acting during a time interval $\Delta t = t_{f}$. On the other hand, this same evolution can be alternatively generated by the Hamiltonian $\hat{\mathcal{H}}(\tau) = \hat{H}(f(\tau))f'(\tau)$ acting during a time interval $\Delta \tau = f^{-1}(t_{f}) - f^{-1}(0)$. A fundamental point to be noticed is that we have freedom of choosing the time-rescaling function $f(\tau)$, and this choice is what determines whether the alternative TR protocol is slower ($\Delta \tau > \Delta t$) or faster ($\Delta \tau < \Delta t$) than the reference driving protocol. In the latter case, the TR evolution would represent a shortcut to the final state. Yet, some remarks must be made on the state of the system and the reference time evolution. First, the initial and final states, $\ket{\psi(t_{f})}$ and $\ket{\psi(0)}$, do not have to be eigenstates of the initial and final Hamiltonians, respectively. That is to say that the TR protocol is state-independent. Second, there are no constraints on the dynamics of the reference protocol, e.g., whether it is adiabatic or not.

As discussed in the introduction, a case of potential interest for quantum control technologies is when the reference evolution produces an adiabatic transformation in the quantum system from $\ket{\psi(0)}$ to $\ket{\psi(t_{f})}$, and we are able to create an equivalent process which is faster. In the present context, if there is a realizable TR protocol satisfying this demand, it would represent a STA. Let us now focus on this case by assuming that the reference process is adiabatic. In these circumstances, the problem of devising a genuine STA
is reduced to finding
an adequate time-rescaling function $f(\tau)$ such that the initial and final Hamiltonians are equal to those of the reference adiabatic process, and, of course, guarantee that $\Delta \tau < \Delta t$. It is easy to see that this task is accomplished if the following four requirements are fulfilled: (i) the initial times must be equal: $f^{-1}(0) = 0$, (ii) the TR protocol must be faster:  $f^{-1}(t_{f})<t_{f}$, (iii) the initial Hamiltonians must be equal:  $\hat{\mathcal{H}}(f^{-1}(0)) = \hat{H}(0)$, and (iv) the final Hamiltonians must be equal: $\hat{\mathcal{H}}(f^{-1}(t_{f}))=\hat{H}(t_{f})$. Assuming that there exist such a TR Hamiltonian satisfying these four requirements, it is important to observe that a STA also takes place even if the $\hat{H}$'s do not commute at different times. In this case we have
\begin{align}
\label{4}
\hat{\mathcal{U}}(t_{f},0) &= \hat{\mathcal{T}}\exp \left\{ -\frac{i}{\hbar} \int_{0}^{t_{f}} \hat{H}(t) dt \right\} \nonumber \\
&= \hat{\mathcal{T}}\exp \left\{ -\frac{i}{\hbar} \int_{f^{-1}(0)}^{f^{-1}(t_{f})} \hat{\mathcal{H}}(\tau) d\tau \right\},
\end{align}
where $\hat{\mathcal{T}}$ denotes the time-ordering operator.

As can be observed, the problem of satisfying the STA requirements lies in the choice of an appropriate time-rescaling function $f(\tau)$ to be used in Eq.~(\ref{3}) (or Eq.~(\ref{4})). The requirements (i) and (ii) are explicit, and finding a function which satisfies both is trivial. On the other hand, properties (iii) and (iv) can be both fulfilled if $f'(f^{-1}(0)) = f'(f^{-1}(t_{f})) = 1$. Thus, any candidate function $f(\tau)$ meeting these criteria can be used in the expression of Eqs.~(\ref{3}) or~(\ref{4}) to turn it into a STA protocol, as an alternative to the reference (adiabatic) evolution. For example, the function
\begin{equation}
\label{5}
f(\tau) = a \tau - \frac{t_{f}}{2 \pi a} (a-1) \sin \left( \frac{2 \pi a}{t_{f}} \tau\right),
\end{equation}
whose inverse function $f^{-1}(\tau)$ cannot be written exactly in terms of standard functions, has the properties $f^{-1}(0) = 0$, $f^{-1}(t_{f}) = t_{f}/a$, $f'(0) = 1$ and $f'(t_{f}/a) = 1$, which are exact as can be easily verified.
These elements qualify this function as an appropriate time-rescaling function for any $a>1$, which we call the {\it time contraction parameter}. In other words, the reference protocol given in Eq.~(\ref{2}) can be realized $a$ times faster, with exactly the same effect, applying the TR protocol of Eq.~(\ref{3}) with $f(\tau)$ given as in Eq.~(\ref{5}) \cite{note}. Therefore, under these conditions, the TR protocol gains the status of STA. We want to call attention to the fact that $f(\tau)$ of Eq.~(\ref{5}) is not unique, so that one can look for many other functions that satisfy the STA requirements. As another example, the polynomial function
\begin{equation}
\label{new}
f(\tau) = \frac{2(a^{2}-a^{3})}{t_{f}^{2}} \tau^{3}+\frac{3(a^{2}-a)}{t_{f}} \tau^{2}+\tau
\end{equation}
is also valid as a time-rescaling function, again with $a$ being the time contraction parameter. In fact, we can easily verify that $f^{-1}(0) = 0$, $f^{-1}(t_{f}) = t_{f}/a$, $f'(0) = 1$ and $f'(t_{f}/a) = 1$.

From a mathematical point of view, the time rescaling method to generate shortcuts, as shown for example in Eq.~(\ref{3}), can also be understood as a joint modification in the passage of time $dt \rightarrow f'(\tau) d \tau$, the time dependence of the original Hamiltonian $\hat{H}(t) \rightarrow \hat{H}(f(\tau))$, and the evolution time $\Delta t \rightarrow \Delta \tau$. This is why we call it ``time-rescaled quantum dynamics''. However, we obviously cannot make time passes at a different rate in practice. Then, the trick was to embody the function $f'(\tau)$ into the Hamiltonian $\hat{H}(t) \rightarrow \hat{H}(f(\tau))f'(\tau)$, and recover the natural passage of time $dt \rightarrow d \tau$. In theory, such a modification in the time dependence of the Hamiltonian is achievable for any controlled quantum dynamics.
To illustrate the present proposal, we address three problems of fundamental and practical interest: the parametric oscillator, the transport of a particle in a harmonic trap, and the spin-1/2 system in a magnetic field.

\section{Parametric oscillator}

The parametric oscillator is described by the one-dimensional Hamiltonian 
\begin{equation}
\label{6}
\hat{H}(t) = \frac{\hat{p}^{2}}{2m}+\frac{1}{2}m \omega^{2} (t) \hat{x}^{2}, 
\end{equation}
where $\hat{x}$ and $\hat{p}$ are the position and momentum operators, respectively. The parameter $m$ is the mass of the oscillator, and $\omega(t)$ is a time-dependent angular frequency. Note that the time dependence of $\hat{H}(t)$ is due only to $\omega(t)$, as seen in Fig. 1. Let us consider that we perform a given protocol during a time interval from $t=0$ to $t =t_{f}$, by varying the frequency of the oscillator under a prescribed scheme, which is ruled by the transformation in Eq.~(\ref{2}). Suppose now that we want to shorten the time duration of the process to last from $\tau = 0$ to $\tau = t_{f}/a$, producing the same final state, by using the TR protocol. In this case, the reference Hamiltonian $\hat{H}$ must be replaced by the TR Hamiltonian       
\begin{equation}
\label{7}
\hat{\mathcal{H}}(\tau) =  f'(\tau)\frac{\hat{p}^{2}}{2m}+\frac{1}{2}m \tilde{\omega}^{2} (\tau) \hat{x}^{2}, 
\end{equation}
with $f(\tau)$ as given by Eq.~(\ref{5}), and the TR frequency obeying $\tilde{\omega}^{2} (\tau) = f'(\tau) \omega^{2}(f(\tau))$. The realization of $\tilde{\omega} (\tau)$ takes place by simply changing the intensity and the time dependence of the fields that generate the harmonic potential.

\begin{figure}[ht]
\centerline{\includegraphics[width=8.3cm]{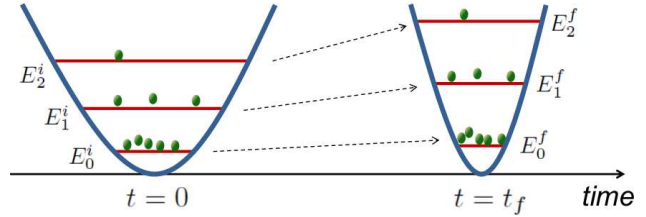}}
\caption{ (Color online) Parametric harmonic oscillator with $\omega (t)$ being an increasing function of time. For the control process to be considered as adiabatic or a STA, the populations of particles in each energy level must be the same both at the beginning and the end of the protocol.}
\label{setup}
\end{figure}

As an example, let us consider the case of a compression stroke of a given quantum heat machine, in which the following boundary conditions are necessary: $\omega (0) = \omega_{0}$, $\omega (t_{f}) = \omega_{f}> \omega_{0}$, $\dot{\omega} (0) = 0$, and $\dot{\omega} (t_{f}) = 0$. The first two conditions characterize the compression, and the last two conditions are necessary for the potential to be static at the beginning and the end of the stroke. With such requirements, we assume now that the angular frequency of the reference process varies according to the relation $\omega (t) = \omega_{0}+(\omega_{f}-\omega_{0}) \sin^{2}(\pi t/2t_{f})$. Therefore, the corresponding angular frequency in the TR protocol becomes
\begin{align}
\label{7.1}
&\tilde{\omega} (\tau) =   \left[ a - (a-1) \cos \left( \frac{2 \pi a}{t_{f}} \tau\right) \right]^{\frac{1}{2}}   \nonumber \\
&  \left\{ \omega_{0}+ (\omega_{f}-\omega_{0}) \sin^{2}\left[ \frac { \pi a} {2 t_{f}} \tau- \left(\frac{a-1}{4a}\right) \sin \left( \frac{2 \pi a}{t_{f}} \tau\right) \right] \right\}. 
\end{align}
Observe that $\omega(0)  = \tilde{\omega} (0) = \omega_{0}$, and $\omega(t_{f})  = \tilde{\omega} (t_{f}/a) = \omega_{f}$, as it should be. In Fig. 2 we display the behavior of the TR angular frequency for some values of the contraction factor.  

\begin{figure}[ht]
\centerline{\includegraphics[width=7.7cm]{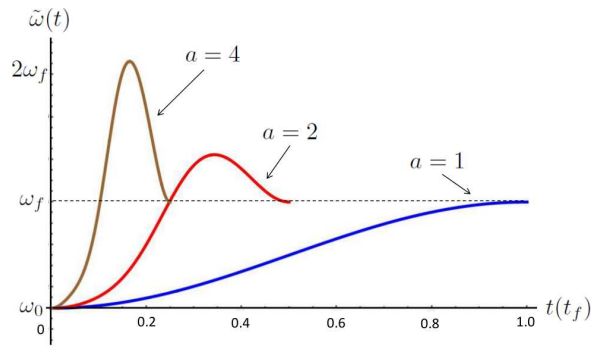}}
\caption{(Color online) Time dependence of the angular frequency of the parametric oscillator for some values of the contraction factor $a$. The $a=1$ curve represents the reference protocol. In all cases we assumed $\omega_{f} = 6 \omega_{0}.$}
\label{setup}
\end{figure}

At the same time, depending on the quantum system we deal with, the modulation of the kinetic energy term with the function $f'(\tau)$ in Eq.~(\ref{7}) may come in different forms. For example, if the trapped particle has a net electric charge $q$, the momentum can be controlled with the application of a time-dependent magnetic field ${\bf{B}}(\tau)$ to cause the transformation $\hat{p} \rightarrow \hat{p} - q {\bf A}(\tau)$, where ${\bf A}(\tau)$ is the time-dependent vector potential satisfying ${\bf B}(\tau) = \nabla \times {\bf A}(\tau)$.  However, as we shall see, for this momentum control to occur it is necessary that the magnetic field be perpendicular to the trapping direction $x$ \cite{landau}.

We now investigate with more detail the magnetic field required to generate the appropriate manipulation of the kinetic energy term. According to the TR protocol, $\hat{H}(t) \rightarrow \hat{H}(f(\tau))f'(\tau)$, the vector potential must satisfy the relation $f'(\tau) \hat{p}^2 = (\hat{p} - q {\bf A}(\tau))^{2}$, which yields ${\bf A}(\tau) = \hat{p}/q \{1 - [f'(\tau)]^{1/2} \} {\bf \hat{x}}$, where ${\bf \hat{x}}$ is the unit vector pointing to the positive $x$ direction. Two important observations have to be made concerning this vector potential. First, it is independent of the coordinates {\it x}, {\it y} and {\it z}. Second, due to gauge invariance, a constant factor added to it at a given instant of time is not physically relevant, but only the way it varies with time. Taking these two facts into consideration, the required potential vector can reduce to the simple form ${\bf A}(\tau) = -B_{0}[f'(\tau)]^{1/2} {\bf \hat{x}}$, where $B_{0}$ is a positive constant. It can easily be verified that the time-dependent magnetic field ${\bf B} (\tau) =  B_{0}[f'(\tau)]^{1/2} {\bf \hat{z}}$ leads to the vector potential  ${\bf A}(\tau) = - B_{0}y[f'(\tau)]^{1/2} {\bf \hat{x}}$, whose dependence on the coordinate $y$ is irrelevant if the charged particle is in fact confined to the $x$ direction. In this form, the field   
\begin{equation}
\label{7.2}
{\bf B} (\tau) = B_{0}  \left[ a - (a-1) \cos \left( \frac{2 \pi a}{t_{f}} \tau\right) \right]^{\frac{1}{2}} {\bf \hat{z}}
\end{equation}
satisfies the required conditions to properly tune the kinetic energy. Observe that this field works independent of the reference protocol, $\omega (t)$, and ${\bf B} (0) = {\bf B} (t_{f}/a) = B_{0}$. Fig. 3 illustrates the time profile of the magnetic field for some values of $a$.  

\begin{figure}[ht]
\centerline{\includegraphics[width=7.7cm]{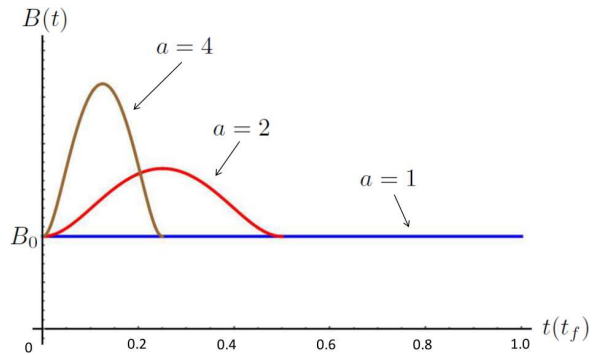}}
\caption{(Color online) Illustration of the time dependence of the magnetic field intensity applied to the parametric oscillator for some values of the contraction factor $a$. The constant field in the $a=1$ case represents the reference protocol.}
\label{setup}
\end{figure}

Now, another important question arises: what is the physical implication of $B_{0}$? To answer this question, we first consider Eq.~(\ref{5}) for $a=1$. In this case, the TR function reduces to $f(\tau) = \tau$, which implies no time rescaling, i.e., the reference protocol. Therefore, if we now look at Eq.~(\ref{7.2}) with $a=1$, we see that this equation is valid if the reference protocol involves the application of a constant magnetic field ${\bf B}_{ref} = B_{0} {\bf \hat{z}}$. That is to say that, if we want that the magnetic field of Eq.~(\ref{7.2}) really provides the kinetic energy manipulation according to our TR approach, the reference protocol must be described by both the time-dependent angular frequency, $\omega(t)$, and the application of ${\bf B}_{ref}$. It signifies that the momentum operator $\hat{p}$ in Eq.~(\ref{7}) is actually given by $\hat{p} = \hat{p}'-q{\bf A}_{ref}$, where $\hat{p}'$ is the momentum of the particle free from any external electromagnetic influence, and ${\bf A}_{ref}$ is the vector potential due to ${\bf B}_{ref}$. In this form, $B_{0}$ is simply the magnitude of the constant magnetic field applied in the reference protocol. In the case of an adiabatic reference protocol, this magnitude must be set so that quantum transitions are prevented.

\section{Transport of a particle by moving a harmonic trap}

We now consider the problem of accelerating the transport of a particle by moving a harmonic trap, which has been previously considered by some authors with different approaches, whose experimental realization is proven to be difficult \cite{odelin}. However, some experiments have been realized transporting ions in a Paul trap by few hundreds of micrometers, preserving the encoded quantum information \cite{bowler,walther}. In order to address this issue, we assume the time-dependent 1D Hamiltonian as given by \cite{odelin0}  
\begin{equation}
\label{7.4}
\hat{H}(t) = \frac{\hat{p}^{2}}{2m}+\frac{1}{2}m \omega^{2} (\hat{x} - x_{0}(t))^{2}, 
\end{equation}
where $x_{0}(t)$ is the ``scalar transport function'', which localizes the position of the minimum of the harmonic potential (see Fig. 4). In experimental applications, it is convenient that the trap starts and ends up at rest in the transport process, so that the scalar function should meet the boundary conditions $x_{0}(0) = 0$, $\dot{x}(0) = 0$, $x_{0}(t_{f}) = d$, and $\dot{x}(t_{f}) = 0$, where $d$ is the distance to be travelled by the trap.

\begin{figure}[ht]
\centerline{\includegraphics[width=8.3cm]{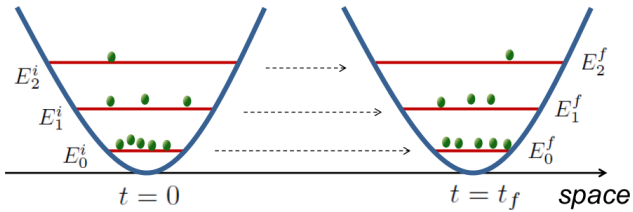}}
\caption{(Color online) Transport of a
particle by moving a harmonic trap. For the driving process to be labeled as adiabatic or a STA, the populations of particles in each energy level must be the same both at the beginning and the end of the time evolution.}
\label{setup}
\end{figure}

Let us assume, for example, that the reference protocol is described by the transport function $x_{0}(t) = d \sin^{2}(\pi t/2t_{f})$, which meets the required boundary conditions. Then, according to the TR method presented here, which says that $\hat{H}(t) \rightarrow \hat{H}(f(\tau))f'(\tau)$, the reference Hamiltonian $\hat{H}$ must be replaced by the TR Hamiltonian       
\begin{equation}
\label{7.5}
\hat{\mathcal{H}}(\tau) =  f'(\tau)\frac{\hat{p}^{2}}{2m}+\frac{1}{2}m \tilde{\omega}^{2} (\hat{x} - \tilde{x}_{0}(\tau))^{2}, 
\end{equation}
with $f(\tau)$ given by the function in Eq.~(\ref{5}), so that the TR angular frequency obeys $\tilde{\omega}^{2} (\tau) = f'(\tau) \omega^{2}$, i.e.,
\begin{equation}
\label{7.6}
\tilde{\omega} (\tau) =   \left[ a - (a-1) \cos \left( \frac{2 \pi a}{t_{f}} \tau\right) \right]^{\frac{1}{2}}\omega. 
\end{equation}
Fig. 5 shows the typical profile of $\tilde{\omega}(\tau)$. Note that $\tilde{\omega}(0) = \tilde{\omega}(t_{f}/a) = \omega $. 

On the other hand, the TR transport function has the form 
\begin{equation}
\label{7.7}
\tilde{x}_{0} (\tau) = d \sin^{2}  \left[  \frac { \pi a} {2 t_{f}} \tau- \left(\frac{a-1}{4a}\right) \sin \left( \frac{2 \pi a}{t_{f}} \tau\right)\right],
\end{equation}
whose typical profile is shown in Fig. 6 for some values of $a$. Observe that $\tilde{x}_{0} (0) = 0$ and $\tilde{x}_{0} (t_{f}/a) = d$, which are the required boundary conditions for a proper TR fast process ($a > 0$). In regards to the kinetic energy term in Eq.~(\ref{7.5}), it is precisely the one found in Eq.~(\ref{7}) for the case of the parametric oscillator. Thus, the application of a magnetic field as described in Eq.~(\ref{7.2}) also works in the transport problem, as long as the particle is charged. Also, the (reference) momentum operator $\hat{p}$ in Eq.~(\ref{7.4}) must be of the type $\hat{p} = \hat{p}'-q{\bf A}_{ref}$, as explained in the previous section. That is, a result of the application of a reference magnetic field 
${\bf B}_{ref} = B_{0} {\bf \hat{z}}$.

\begin{figure}[ht]
\centerline{\includegraphics[width=7.7cm]{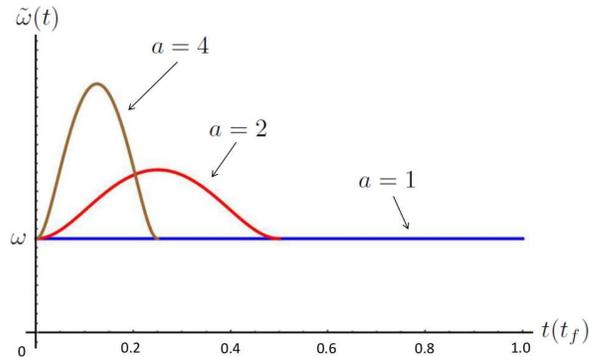}}
\caption{(Color online) Angular frequency of the harmonic trap as a function of time for some values of the contraction factor $a$. The constant $\tilde{\omega}$ curve ($a=1$) corresponds to the reference process, in which the trap does not change in shape.}
\label{setup}
\end{figure}

\begin{figure}[ht]
\centerline{\includegraphics[width=7.7cm]{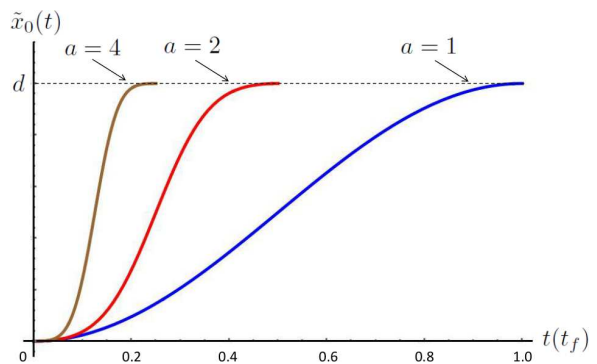}}
\caption{(Color online) Time dependence of the transport function of the harmonic trap for some values of the contraction factor $a$. The reference transport process is represented by the $a=1$ curve. The parameter $d$ is the total distance travelled by the trap in the protocols.}
\label{setup}
\end{figure}

\section{Spin-1/2 system in a magnetic field}

Now we turn to the study of a spin-1/2 particle in a time-varying magnetic field ${\bf B}(t)$. In this case, the Hamiltonian is given by \cite{sakurai}
\begin{equation}
\label{8}
\hat{H}(t) = \gamma {\bf B}(t) \cdot\ \hat {{\bf S}} , 
\end{equation}
where $\gamma$ is the gyromagnetic ratio and $\hat {{\bf S}}$ is the vector spin operator of the particle, i.e., $\hat {{\bf S}} = \hbar/2(\hat{\sigma}_{x},\hat{\sigma}_{y},\hat{\sigma}_{z})$, involving the Pauli matrices. 
Assuming that the Hamiltonian of Eq.~(\ref{8}) generates the reference process, we have that the TR evolution operator is given as in Eq.~(\ref{4}), with the time-rescaling function according to Eq.~(\ref{5}). That is,
\begin{align}
\label{9}
\hat{\mathcal{U}}(t_{f},0) &= \hat{\mathcal{T}}\exp \left\{ -\frac{i}{\hbar} \int_{0}^{t_{f}} \hat{H}(t) dt \right\} \nonumber \\
&= \hat{\mathcal{T}}\exp \left\{ -\frac{i}{\hbar} \int_{0}^{t_{f}/a} \hat{\mathcal{H}}(\tau) d\tau \right\},
\end{align}
with 
\begin{align}
\label{10}
\hat{\mathcal{H}}(\tau) = &\gamma  \left[ a - (a-1) cos \left( \frac{2 \pi a}{t_{f}} \tau\right) \right]  \nonumber \\
& {\bf B} \left( a \tau - \frac{t_{f}}{2 \pi a} (a-1) sin \left( \frac{2 \pi a}{t_{f}} \tau\right) \right) \cdot \hat {{\bf S}}. 
\end{align}

As a simple demonstration, we now consider the familiar case of a spin-1/2 particle in a constant magnetic field oriented along the $z$-axis, ${\bf B}(t) = B_{0} \hat{{\bf z}}$, as the reference protocol. The Hamiltonian is given simply by $\hat{H}(t) = \hat{H} = \Omega \hat{S}_{z}$, with $\Omega = \gamma B_{0}$. In this form, the evolution operator becomes $\hat{\mathcal{U}}(t_{f},0) = exp(-i \Omega \hat{S}_{z} t_{f}/\hbar)$. We will also assume that at $t=0$ the particle is in the state $\ket{\psi(0)} = \ket{S_{x},+} = 1/\sqrt{2}(\ket{+}+\ket{-})$, where $\hat{S}_{z} \ket{\pm} = \pm \hbar/2 \ket{\pm}$. Accordingly, we have $\hat{\mathcal{U}}(t_{f},0) = e^{-i \Omega t_{f}/2} \ket{+} \bra{+} + e^{i \Omega t_{f}/2} \ket{-} \bra{-}$, which provides that the state of the system after a time $t_{f} = \pi/\Omega$ is $\ket{\psi( \pi/\Omega)} = \hat{\mathcal{U}}(\pi/\Omega,0) \ket{\psi(0)} = -i/\sqrt{2}(\ket{+}-\ket{-}) = \ket{S_{x},-}$, up to a global phase factor $-i$. Overall, we observe that $\Delta t = \pi/\Omega$ is the shortest time interval for which the constant magnetic field ${\bf B} = B_{0} \hat{{\bf z}}$ causes a spin flip in the $x$-direction. The uncertainty in energy of the initial state $\ket{\psi(0)}$ can be found to be $\Delta E = \sqrt{\langle \hat{H}^{2} \rangle - \langle \hat{H} \rangle ^{2}} = \hbar \Omega/ 2$. Hence, this configuration satisfy the relation $\Delta t  \Delta E = \hbar \pi/2$, which is the limit of the Mandelstam-Tamm bound \cite{mand}, i.e., the quantum speed limit \cite{deff2}.  

Now we investigate the effect of the TR protocol obtained from the reference process above. In this regard, for us to achieve an identical spin flip with the same initial and final Hamiltonians, in a shorter time interval, we should apply the Hamiltonian \begin{equation}
\label{11}
\hat{\mathcal{H}}(\tau) = 
 \left[ a - (a-1) \cos \left( \frac{2 \pi a}{t_{f}} \tau \right) \right] \Omega  \hat{S}_{z}
\end{equation}
between $\tau = 0$ and $\tau = t_{f}/a = \pi/\Omega a$, with the contraction factor $a>1$. Indeed, if we have again the initial state as $\ket{\psi(0)} = \ket{S_{x},+}$, the TR evolution generated by $\hat{\mathcal{H}}(\tau)$ produces

\begin{align}
\label{12}
\ket{\psi(\pi/\Omega a)} &= \hat{\mathcal{U}}(\pi/\Omega a,0) \ket{\psi(0)} \nonumber \\ &= \exp \left\{ -\frac{i}{\hbar} \int_{0}^{\pi/ \Omega a} \hat{\mathcal{H}}(\tau) d\tau \right\} \ket{S_{x},+} \nonumber \\
&= \exp \left\{ -\frac{i \hat{S}_{z} }{ \hbar} \pi   \right\} \ket{S_{x},+} = \ket{S_{x},-},
\end{align}
up to the same global phase factor $-i$ of the original protocol. As desired, the TR protocol had the same effect of the reference one, with the same final Hamiltonian. Similar to the previous examples, the reference protocol with the Hamiltonian $\hat{H}$ does not have to be necessarily adiabatic. In such case, as already mentioned, the TR process generated by $\hat{\mathcal{H}}(\tau)$ satisfies all STA requirements. About the time energy uncertainty relation, we have now that $\Delta t \Delta E = \hbar \pi / 2a$. This result is not a violation of the Mandelstam-Tamm limit because this bound is not valid for driven dynamics, i.e., parametrically varying Hamiltonians \cite{deff2,campa}. For the sake of comparison, in Fig. 7 we show the time dependence of the magnetic fields for the reference and TR processes with different values of $a$.   
\begin{figure}[ht]
\centerline{\includegraphics[width=7.7cm]{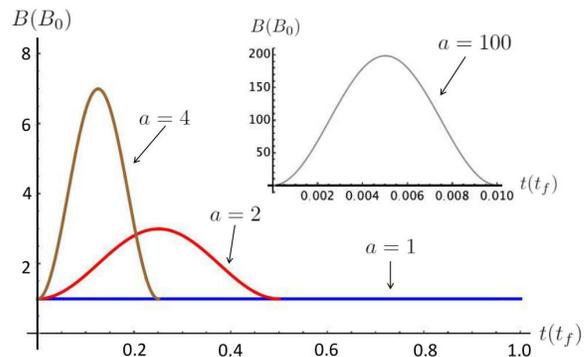}}
\caption{(Color online) Time dependence of the magnetic field intensity for some values of the contraction factor $a$. We observe that the shorter the time interval of the TR protocol, the higher the intensities involved. As required in the STA criteria, all fields have the same intensity $B_{0}$ of the reference ($a=1$) protocol at the initial and final times.}
\label{setup}
\end{figure}

Additional important information can be obtained from the curves shown in Fig. 7 with respect to the time rescaling method. For example, the shorter the time demanded to realize a given dynamics, the higher the intensity of the field required. This characteristic is not exclusive to the present example, or to a restrict class of quantum systems. In fact, the integrals composing the argument of the exponential operators in Eqs.~(\ref{2}) and~(\ref{3}), $\int_{0}^{t_{f}} \hat{H}(t) dt$ and $\int_{f^{-1}(0)}^{f^{-1}(t_{f})} \hat{\mathcal{H}}(\tau) d\tau$, respectively, must be equal if we want that the reference and TR processes produce the same resulting effect. However, these definite integrals can be geometrically understood as the area under the energy versus time curve, which means that a shorter TR process demands more energy, i.e, the external fields responsible for generating the dynamics must be more and more intense. Therefore, in practice, the achievement of very short TR processes is limited both by the intensity of the fields employed, and the fidelity of its actual time evolution to the prescribed protocol within the desired short time interval.           

\section{Comparative analysis with other STA methods}

At this stage, it is interesting to compare our TR quantum dynamics proposal with other existing STA methods \cite{odelin}. First, we notice that our preliminary idea of ``accelerating'' a given quantum process is similar to that of the FF dynamics \cite{masuda,masuda2}. However, in that proposal the objective is to derive an alternative potential from a reference one in which the wavefunction of the system have its dynamics accelerated at some rate $\alpha$, called magnification factor. The focus of that method is mostly on the dynamics of spatial wavefunctions, which faces a problem when the Hamiltonian has a kinetic energy term, requiring a change in the mass of the particle. Here, when solving the problem of the parametric oscillator, we also faced a similar obstacle. Nevertheless, we showed that this can be circumvented by manipulating the momentum with an external time-dependent magnetic field, whereas the FF approach deals with it by keeping the kinetic energy term unchanged, modifying only the potential, unavoidably making it nonlinear, and externally controlling the phase of the wavefunction. 

In general, both methods seem to be more difficult to realize experimentally for continuous Hamiltonians, but the present one has been shown to be rather simpler for discrete ones. Moreover, another advantage in using the TR approach is that, contrary to the FF method \cite{masuda3}, it is state-independent, as could be noticed in the examples above.  This can also be seen from the fact that in developing the TR protocol, we only manipulated the evolution operator, leaving the initial ket state unchanged (see the discussion after Eq.~(\ref{3})).  For the sake of comparison, we refer to the application of the FF protocol to some of the cases studied here: the parametric oscillator \cite{SMG}, and the spin-1/2 in a magnetic field \cite{taka}. In both situations, the auxiliary FF potential that generates the STA evolution depends on the initial state of the quantum system.                

With respect to the CD method, as mentioned in the introduction, one of the difficulties is that one needs to know the instantaneous eigenkets of the reference Hamiltonian $\hat{H}_{0}(t)$,   $\ket{n_{t}}$, a problem that is not found here. Another point is that the auxiliary Hamiltonian $\hat{H}_{1}(t)$ are in many cases of interest computable analytically, even for continuous spectra, but the experimental realization is problematic \cite{torr}. Comparatively, a further point which must be mentioned is that, contrary to CD processes, our TR protocol is not transitionless. In fact, if we observe Eqs.~(\ref{2}) and~(\ref{3}), we see that the reference and TR evolution operators are only equivalent for those specific limits of integration, i.e., they differ from each other at intermediate times.  It means that if the reference protocol is adiabatic, i.e., the physical system remains in its instantaneous eigenstate along the whole evolution, the corresponding TR process will only provide this effect with certainty at the final time.

\section{Conclusion}

We have proposed a state-independent mechanism to speed up the evolution of a quantum state by rescaling the time dependence of a reference evolution protocol. In comparison with the reference protocol, the time duration of the proposed TR process can be shortened by an arbitrarily large contraction factor. Moreover, the initial and final Hamiltonians are preserved, and no information about the spectrum of the system is needed. For the case in which the reference protocol is adiabatic, it is shown that the TR protocol works as a shortcut to adiabaticity, which has been proven to have a number of practical applications in the quantum control of several systems. To illustrate the present findings, we discussed our approach under the perspective of the parametric oscillator, the transport of a particle in a harmonic trap, and the spin-1/2 particle in a magnetic field. Overall, we believe that the present speed up operation has potential applications in many quantum technologies as finite-time quantum thermodynamics and many-body state engineering.  

\section*{ACKNOWLEDGMENTS}
        
The author acknowledges financial support from the Brazilian funding agencies CAPES/Finance Code 001, and CNPq, Grant Number 309292/2016-6.


\begin{thebibliography}{7}
\bibitem{note} The same argument is valid relating the reference and TR processes in Eq.~(\ref{4}).
\end{thebibliography}
\end{document}